\documentclass[fullpaper]{nldl}

\renewcommand{\DOI}{12345}

\usepackage[utf8]{inputenc}
\usepackage{url}
\usepackage{graphicx}
\usepackage{authblk}
\usepackage{verbatim}
\usepackage{amsfonts, amsmath, amssymb}
\usepackage{algorithm}
\usepackage{algorithmic}
\usepackage{lipsum}
\usepackage{booktabs}
\usepackage{multirow}
\usepackage[inline]{enumitem}
\usepackage{hyperref}
\usepackage{balance}
\usepackage[subtle]{savetrees}

\title{Continuous Metric Learning For Transferable Speech Emotion Recognition and Embedding Across Low-resource Languages}
\author[1]{Sneha Das
}
\author[1]{Nicklas~Leander~Lund
}
\author[2]{Nicole Nadine Lønfeldt
}
\author[2,3]{Anne Katrine Pagsberg
}
\author[1]{Line H. Clemmensen
}
\affil[1]{Department of Applied Mathematics and Computer Science, Technical University of Denmark}
\affil[2]{Child and Adolescent Mental Health Center, Copenhagen University Hospital, Capital Region}
\affil[3]{Faculty of Health, Department of Clinical Medicine, Copenhagen University}
\date{\vspace{-5ex}}

\begin{document}
\nldlmaketitle

\begin{abstract}  
Speech emotion recognition~(SER) refers to the technique of inferring the emotional state of an individual from speech signals. SERs continue to garner interest due to their wide applicability. Although the domain is mainly founded on signal processing, machine learning, and deep learning, generalizing over languages continues to remain a challenge. However, developing generalizable and transferable models are critical due to a lack of sufficient resources in terms of data and labels for languages beyond the most commonly spoken ones. To improve performance over languages, we propose a denoising autoencoder with semi-supervision using a continuous metric loss based on either activation or valence. The novelty of this work lies in our proposal of continuous metric learning, which is among the first proposals on the topic to the best of our knowledge. Furthermore, to address the lack of activation and valence labels in the transfer datasets, we annotate the signal samples with activation and valence levels corresponding to a dimensional model of emotions, which were then used to evaluate the quality of the embedding over the transfer datasets\footnote{The labels are available at:\url{https://bit.ly/3rg6VsA}}. We show that the proposed semi-supervised model consistently outperforms the baseline unsupervised method, which is a conventional denoising autoencoder, in terms of emotion classification accuracy as well as correlation with respect to the dimensional variables. Further evaluation of classification accuracy with respect to the reference, a BERT based speech representation model, shows that the proposed method is comparable to the reference method in classifying specific emotion classes at a much lower complexity.
\end{abstract}
\begin{figure}[!t]
\centering
\includegraphics[width=0.9\columnwidth]{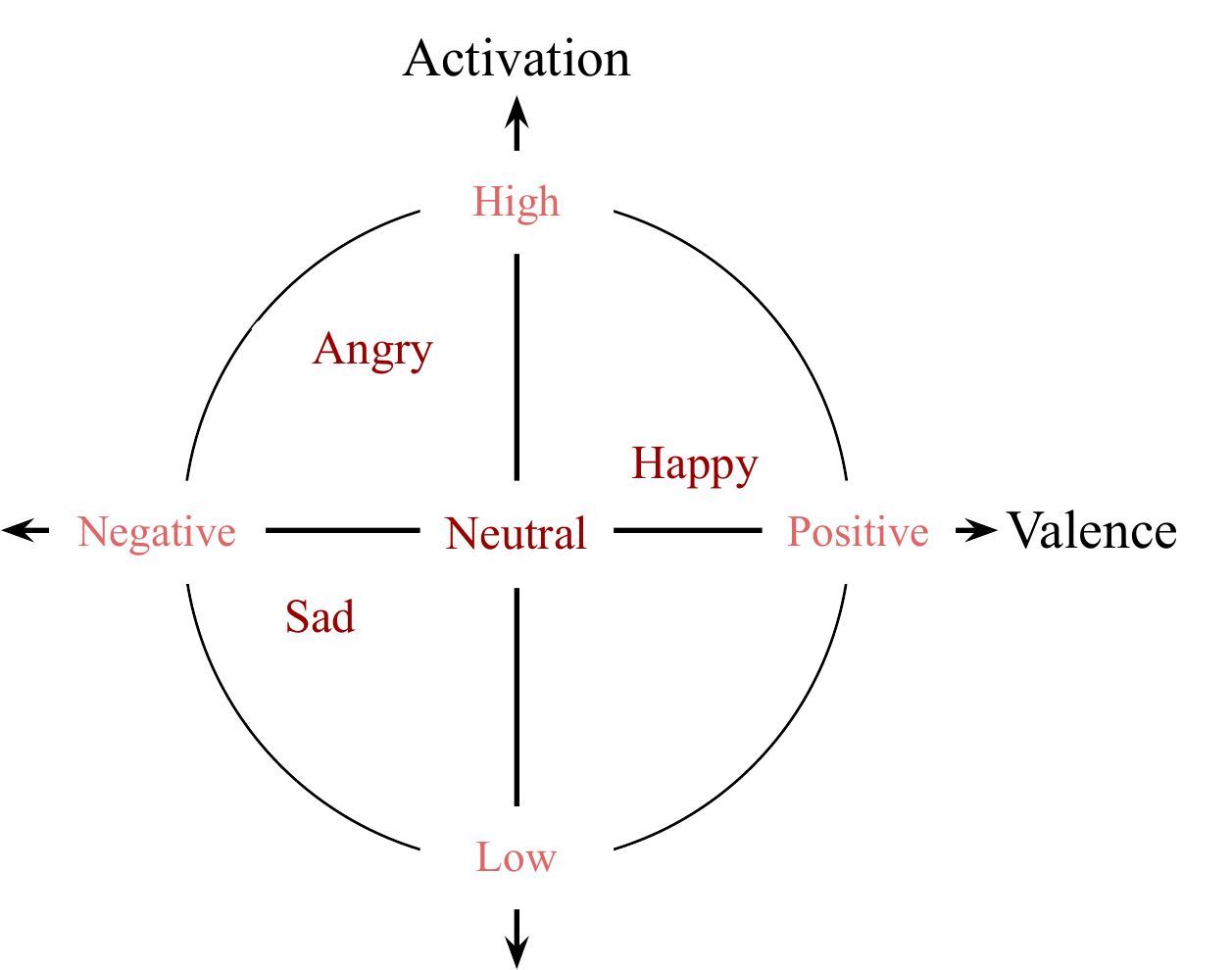}
\caption{Illustration of the circumplex dimensional model of emotion~\cite{russell1980circumplex}.}
\label{fig:circumplex}
\end{figure}
\section{Introduction}
\label{sec:intro}
 Speech emotion recognition~(SER) is the process of inferring the emotional state from speech signals. The domain has found applicability in diverse fields, in healthcare in the detection of disorders, risk assessment in criminal justice system, monitoring the attentiveness of students in school, etc. 

Methods for SER have evolved from solely signal processing and conventional machine learning methods to using more advanced deep learning. Notable methods employed the use of hidden Markov models and Gaussian mixture models for emotion classification. Support vector machines have been a widely used tool for SER, either as a standalone tool or in tandem with other methods to improve classification~\cite{schuller2004speech}. Furthermore, advanced machine learning methods including deep learning models have recently shown promise for SER. This includes the use of CNNs, RNNs and LSTMs ~\cite{wollmer2010combining}.     

An enduring struggle of SER models is their ability to generalize over languages. This is addressed by using supervision when the languages are supported with sufficient resources, both in data and labels. However, supervised learning is inapplicable to languages beyond the commonly spoken ones, due to insufficient resources in terms of small data sets and few-to-no labels. Therefore, to cater to low-resource languages, it is necessary to develop methods that generalize better over languages. The associated challenge is the subjectivity in emotion perception and classification. For instance, the perception of speech signals demonstrating a neutral emotion can vary between languages, owing to phonetic and cultural differences. Despite this, most models use class labels to train models, instead of employing a more universal and continuous metric like activation and valence from the dimensional model of emotions~\cite{russell1980circumplex}.

Latent representation methods, like the autoencoder~(AE), can compress the input features to a smaller and ideally more target relevant latent embedding and are commonly employed in various applications, for instance biosignal processing, computer-vision, and speech-processing~\cite{yildirim2018efficient}. Furthermore, these methods are relatively more interpretable. This is a favourable characteristic as the decision making process of the model is more transparent thereby enabling smoother deployment of such systems in medical and clinical setup. Therefore, latent representation methods can also be useful in the modelling of emotions from speech signals, such that the models retain only the emotion-relevant paralinguistic content from the speech signal. This may also lead to better knowledge transfer between data sets by transferring only generic emotion representations to unseen languages and corpora and not the syntactical variations between languages. Transferable emotion representations aid in addressing the acute label shortage for SER tasks in under-resourced languages. 

In this paper, we use the denoising autoencoder~(DAE) to obtain a highly compressed emotion embedding that is more consistent over different languages. Additionally, we strive to keep the system relatively simple and interpretable as the target application of this work is clinical psychiatry. While a conventional DAE will learn a compressed latent space representative of the input features, it is not necessary that the model learns to discriminate between the desired emotions or emotion representation. Therefore, we propose to use semi-supervision to direct the DAE to learn factors relevant to emotion discrimination like activation and valence, wherein the semi-supervision is provided by the activation and valence labels. We address the following questions in the process: \begin{enumerate*}\item How to provide semi-supervision to preserve the label distances between data samples in the latent space, where the labels span a continuous space. In contrast, most works on metric learning, like~triplet loss and contrastive loss are discrete in nature. \item How to validate the embedding quality in the absence of dimensional labels~(activation and valence) for the transfer datasets considered in this paper. \end{enumerate*} Therefore, the contributions of this work are: \begin{enumerate*} \item We propose a method for continuous metric learning to order the samples in the latent space. To the best of our knowledge, this is one of the few methods addressing continuous metric learning. \item We annotate the transfer datasets with the activation and valence labels that we use to validate the effectiveness of the proposed methods over languages. The labels will be shared for research purposes. 
\end{enumerate*}

\begin{figure}[!t]
\centering
\includegraphics[width=0.9\columnwidth]{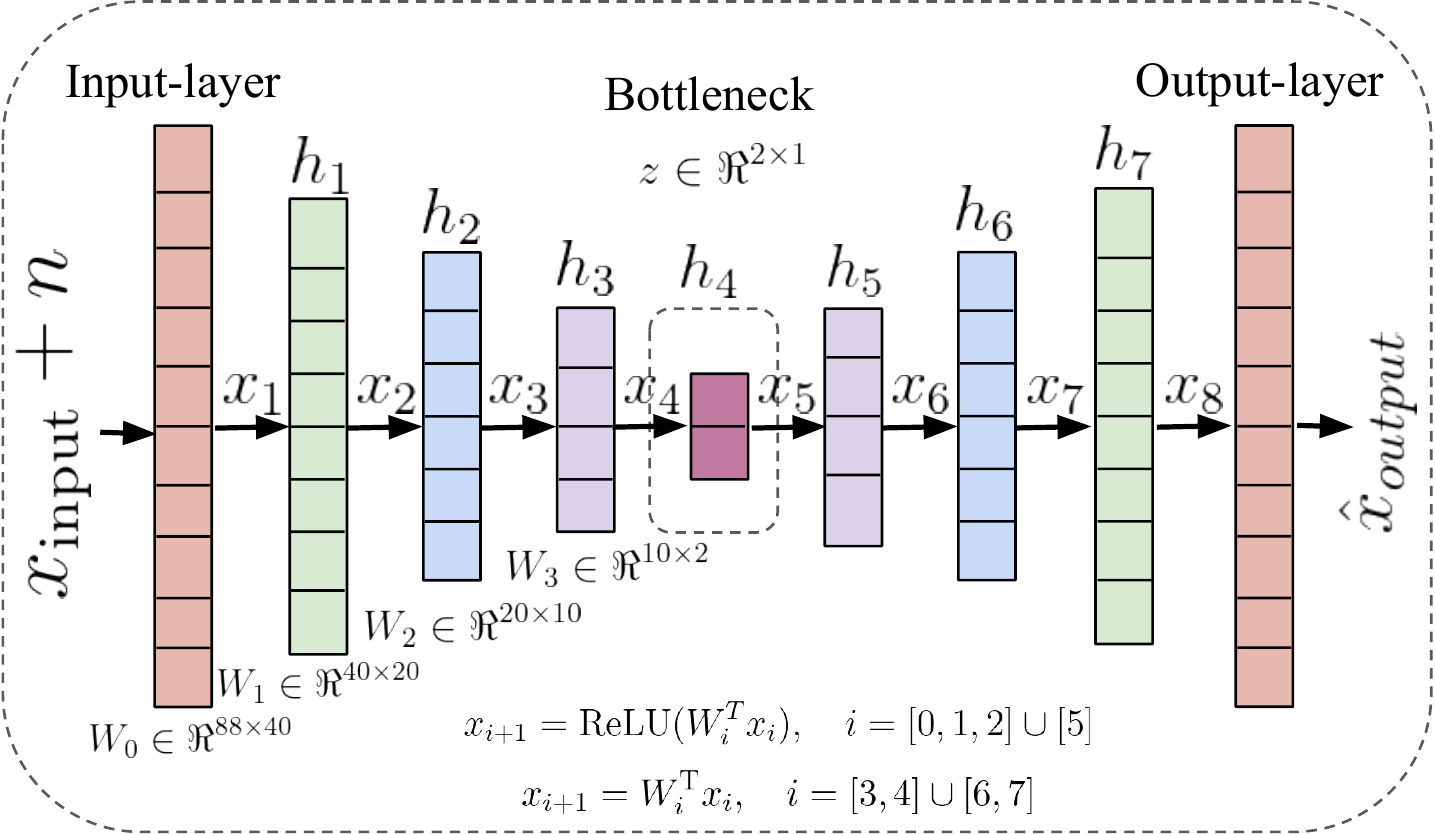}
\caption{Illustration of the (denoising) autoencoder architecture.}
\label{fig:archi}
\end{figure}

\section{Relation To Prior Work}
\label{sec:prior}

 \paragraph*{Autoencoders and variants:}DAE was one of the earliest deep learning based unsupervised learning techniques for SER~\cite{xia2013using}. This was followed by the use of sparse AE for feature transfer~\cite{deng2013sparse} and for SER on spontaneous data sets~\cite{dissanayake2020speech}. Furthermore, end-to-end representation learning for affect recognition from speech was proposed and showed performance comparable to existing methods~\cite{ghosh2016representation}. In recent years, techniques like variational and adversarial AEs and adversarial variational Bayes have been exploited to learn the latent representations of speech emotions with input features ranging from the raw signals to hand crafted features~\cite{latif2018variational, parthasarathy2019improving, eskimez2018unsupervised, neumann2019improving}. 

\paragraph*{Metric learning in SER:}In a recent paper on SER, the authors proposed a convolutional autoencoder that employs a pre-trained autoencoder and a convolutional neural network, and a triplet loss in the autoencoder is used for metric learning~\cite{gao2021metric}. Further on, a contrastive loss was used for metric learning in a Siamese network~\cite{lian2018speech}. On similar lines, a class specific triplet-loss based LSTM neural network was proposed for SER~\cite{huang2018speech}. A combination of the centre-loss, that clusters members of a class together, and the cross-entropy loss was investigated to better cluster features~\cite{dai2019learning, mocanu2021utterance}. The authors proposed multiple f-similarity preserving losses for metric learning using soft labels and tested it on classification~\cite{zhang2019f}. 

\section{Methodology}
\paragraph*{Dimensional Model of Emotion:}
We consider the circumplex model of emotion, wherein Russell et al proposed that emotions can be represented in a circular space, wherein the x-axis corresponds to valence, and y-axis corresponds to the activation~\cite{russell1980circumplex}, as illustrated in Fig.~\ref{fig:circumplex}. Activation refers to how arousing an emotion is and valence refers to the level of positivity in the emotion.

\subsection{Denoising autoencoder with continuous metric learning}
\label{sec:DAE_ss}

\begin{figure*}[!t]
\centering
\includegraphics[width=0.99\textwidth]{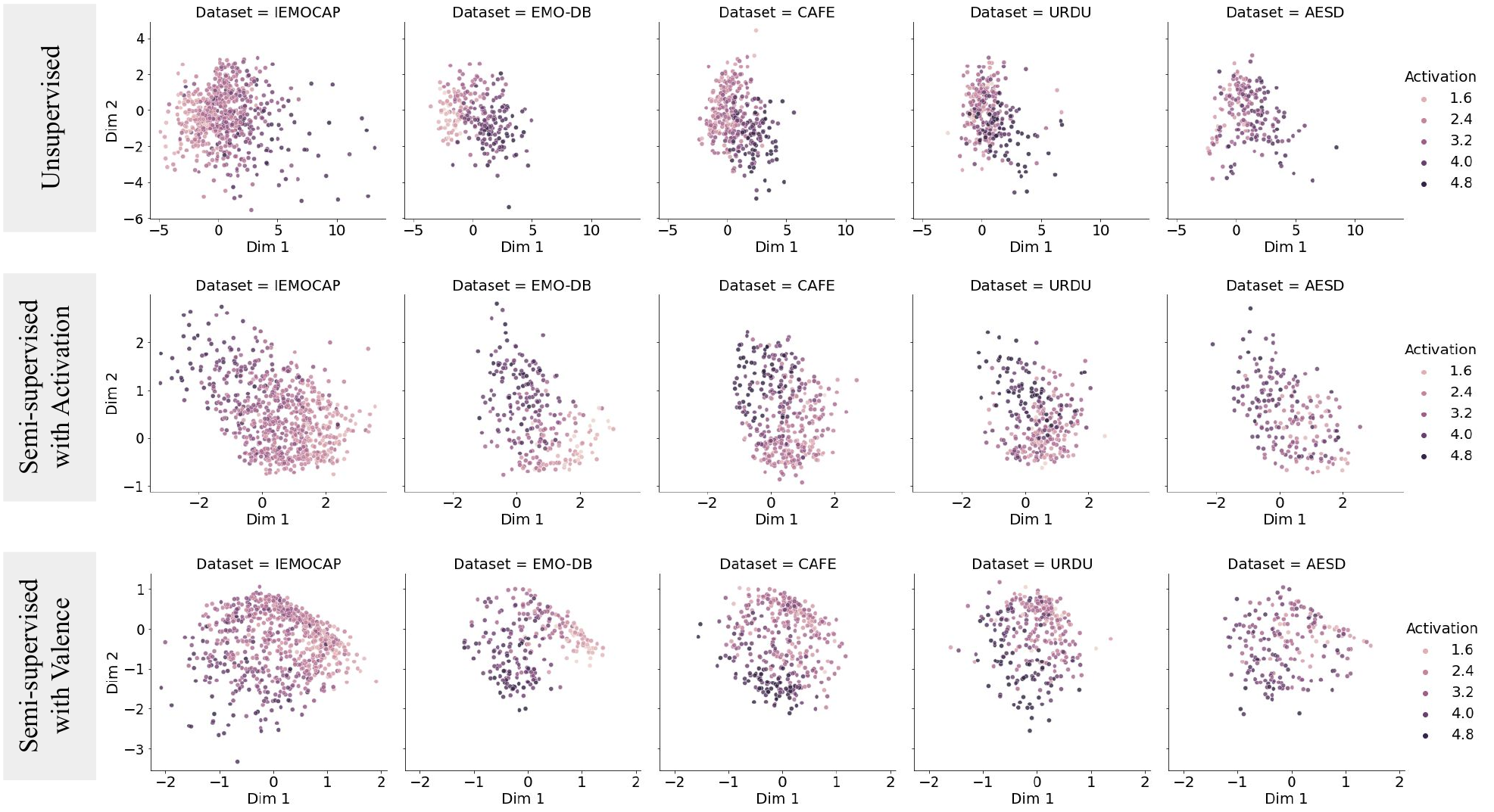}
\caption{Latent embedding for DAE-unsupervised~(Top), DAE-metric-act~(Middle) and DAE-metric-val~(Lower) over five datasets, color-coded by activation levels. The models have been trained on IEMOCAP only.}

\label{fig:latent_emb}
\end{figure*}

We represent the learning function in a denoising autoencoder as: 
\begin{equation}
\begin{aligned}
\text{arg}\min_{f_\theta, g_\phi} \quad \mathcal{L}_{\text{rec}}=\mathbb{E}\lVert \mathbf{x}-g_\phi(f_\theta(\mathbf{x_n})) \rVert_2^2, 
\end{aligned}
\label{eq:1}
\end{equation}
where $\mathbf{x}$ and $\mathbf{x_n}$ are the clean and partially noisy feature vectors, $f_\theta$ represents the encoder, $g_\phi$ the decoder, and $\mathcal{L}_{\text{rec}}$ is the reconstruction loss~\cite{vincent2008extracting}. It is known to be more robust than the AE, since it is designed to learn a subspace of the clean feature vectors from a noisy input, therefore learning the characteristics of the desired input.

Let $\mathbf{z}\in\mathbb{R}^{b \times 2}$ be the latent space embedding, and $\mathbf{l}\in\mathbb{R}^{b \times 1}$ be the label we are modelling. Then $\mathbf{z_d}=d(z_i, z_{i+1})$, such that $\mathbf{z_d} \in \mathbb{R}^{(b-1)\times1}$, where $d(a_1, a_2)$ is the Euclidean distance between the vectors $a_1$ and $a_2
$, and $\mathbf{z_d}$ is the distance between two latent samples. Similarly, $\mathbf{l_d}=d(l_i, l_{i+1})\in \mathbb{R}^{(b-1)\times1}$, is the distance between the labels of the data samples, and $i=1, 2, \dots, (b-1)$.
We assume that $\mathbf{z_d}$ is a linear function of $\mathbf{l_d}$ such that $\mathbf{\hat{z}_d} = p\mathbf{l_d}$; we obtain the optimum $p$ by minimizing the squared error $\lVert \mathbf{z_d}-\mathbf{\hat{z}_d}\ \rVert_2^2$, which yields:
\begin{equation}
    p = (\mathbf{l_d}^T\mathbf{l_d})^{-1}\mathbf{l_d}^T\mathbf{z_d}
\end{equation}

As motivated in the Sec.~\ref{sec:intro}, our goal is to obtain a latent space through training, wherein the distance between the embedding of two samples is close to the distance between the labels of the corresponding samples. We can achieve this by \begin{enumerate*} \item obtaining a slope that is close to one, and \item minimizing the residual between $\mathbf{z_d}$ and  $\mathbf{\hat{z}_d}$ \end{enumerate*}. Therefore, the loss factor corresponding to the slope between $\mathbf{z_d}$ and $\mathbf{l_d}$ is given as:
\begin{equation}
    \mathcal{L}_{\text{sl}} = \left\| \frac{\mathbf{\hat{z}_d}(a_1)-\mathbf{\hat{z}_d}(a_2)}{\mathbf{{l}_d}(a_1)-\mathbf{{l}_d}(a_2)} -1\right\|_2, 
\end{equation}
where $a_1, a_2$ are two arbitrary data instances. 
Furthermore, the residual component is the standard mean squared error: 
\begin{equation}
    \mathcal{L}_{\text{res}} = \mathbb{E}\lVert \mathbf{z_d}-\mathbf{\hat{z}_d} \rVert_2^2. 
\end{equation}
Therefore, our proposed total metric loss is composed of:  
\begin{equation}
\label{eq:met}
    \mathcal{L}_{\text{met}} = \mathcal{L}_{\text{res}} + \mathcal{L}_{\text{sl}}
\end{equation}

The final equation for the DAE with metric learning is given as: 
\begin{equation}
\begin{aligned}
\text{arg}\min_{f_\theta, g_\phi} \quad \mathcal{L}_{\text{rec}}+\mathcal{L}_{\text{met}}, 
\end{aligned}
\label{eq:2}
\end{equation}
where $\mathcal{L}_{\text{rec}}$ has the same formulation as in Eq.~\ref{eq:1}

\subsection{Pair-wise distance preservation}
\label{sec:OLS}

\begin{table}[!tbh]

    \resizebox{.99\columnwidth}{!}{

    \begin{centering}
\begin{tabular}{lrr}
\toprule
       Method &  $R^2$-Act~($\mu \pm \sigma$) &  $R^2$-Val~($\mu \pm \sigma$) \\
\midrule
 Unsupervised &       0.11$\pm$0.06 &       0.03$\pm$0.02 \\
   Metric-act &       {$\mathbf{0.21\pm 0.05}$} &      $\mathbf{0.06\pm0.02}$\\
   Metric-val &       0.12$\pm$ 0.05 &      0.05$\pm$0.02 \\
\bottomrule
\end{tabular}
\end{centering}
    }
        \caption{Adjusted squared correlation coefficient presenting the linear dependence of $\mathbf{z_d}$ on $\mathbf{l_d}$ for the three models. Mean and standard deviation over five folds are presented.}
    \label{table:1}

\end{table}
To obtain an insight on the effectiveness of the proposed metric loss, we study the correlation between $\mathbf{z_d}$ and $\mathbf{l_d}$ by using ordinary least squares~(OLS) to model $\mathbf{l_d}$ as a function of $\mathbf{z_d}$ as follows: $\mathbf{l_d} = c+\beta_1\mathbf{z_{d_1}}+\beta_2\mathbf{z_{d_2}}$. The assumption of a linear relation between the pair-wise point distances in the latent space and the labels is motivated from our proposed metric loss. The $R^2$ adjusted is presented in Table~\ref{table:1}.

\section{Experimental setup}
In this section, we describe the dataset and the input feature space, the architecture of the models, the pre-processing steps followed by the description of the methods of comparison and our evaluation setup. 

\paragraph*{Datasets and input features:}
\label{sec:data}
IEMOCAP, an audio-visual affect data set, is used to train and validate the models~\cite{busso2008iemocap}. The data set comprises of annotations representing both the categorical and dimensional emotional model~\cite{bakker2014pleasure}. The models are trained with data from the emotional categories {\it neutral~(N), sad~(S), happy~(H), angry~(A)}. We use the extended Geneva minimalistic acoustic parameter set~(eGeMAPS)~\cite{eyben2015geneva}, specifically the functionals of lower-level features. Each speech sample yields a feature vector comprising of 88 features and we use the OpenSmile toolkit to extract the features~\cite{eyben2010opensmile}. To study how the latent representations are transferred between corpora and languages, we test classification accuracy over the following {\it transfer datasets} :  \begin{enumerate*} \item the Surrey Audio-Visual Expresses Emotion~(SAVEE) database that is primarily English and consists of male speakers only, \item the Berlin Database of Emotional Speech~(Emo-DB) recorded in German and, \item the Canadian French Emotional~(CAFE) speech database comprising of French audio samples~\cite{jackson2014surrey, burkhardt2005database, gournay2018canadian}, \item the URDU-dataset consisting of Urdu speech, and \item the Acted Emotional Speech Dynamic~(AESD) Database comprising of Greek speech~\cite{latif2018cross, vryzas2018speech}. Note that AESD does not comprise of speech samples from the neutral class, whereby evaluation includes AESD speech samples from three and two classes only.\end{enumerate*} For correlation analysis, due to the lack of activation and valence labels, we annotate the following datasets: \begin{enumerate*} \item EMO-DB, \item URDU \item AESD and \item CAFE \end{enumerate*}. These labels are then used to test the baseline DAE-unsupervised, and the proposed DAE-metric-act and DAE-metric-val models.
\vspace{-0.3cm}
\paragraph*{System architecture:}
\label{sec:architect}
Past works using AEs and variants have mostly addressed novel network architectures for better classification accuracy~\cite{dissanayake2020speech, latif2018variational, neumann2019improving, parthasarathy2019improving, xia2013using}. However, since the focus of this work is to investigate the potential of metric learning in obtaining a more transferable embedding space for emotion recognition, we employ a simple architecture for the DAE baseline with performance that is comparable to existing methods. The proposed methods share the same architecture as the baseline model, illustrated in Fig.~\ref{fig:archi}. The size of the input features is 88 and we compress the latent space to two dimensions. As described above, the input feature vector is comprised of the descriptive statistics of each speech signal, without temporal correlation. Therefore, in the encoder and decoder we employ fully connected NNs, instead of convolutional or recurrent NNs. The fully connected layers are followed by rectified linear units~(ReLU) to incorporate non-linearity within the model. 
\vspace{-0.3cm}
\paragraph*{Preprocessing:}
\label{sec:pp}
Prior to using the data sets for training and testing, we remove the outliers by computing the z-score and eliminating the data samples that have a z-score, $-10>z>10$. We chose a threshold of 10 instead of the standard value of 3 because the goal of this work is to understand the behavior of the models for both typical and atypical rendition of emotions in speech. Therefore, we only remove the extreme outliers. For run-time evaluation over the transfer datasets, we employ statistics from 20\% of the transfer data set to standardize the remaining 80\%. 
\vspace{-0.3cm}
\paragraph*{Proposed methods:}
\label{sec:methods}
We train and validate the following models: \begin{enumerate*}\item DAE-Unsupervised is used as a baseline, \item DAE-Metric-Act that utilizes the proposed semi-supervision via the activation labels from Sec.~\ref{sec:DAE_ss}, and \item DAE-Metric-Val utilizes the proposed semi-supervision via the valence labels~(Sec.~\ref{sec:DAE_ss}). \end{enumerate*}
The input features to the DAE is corrupted by a noise component, $x_{\text{input}}=x_{\text{true}}+N$ and $N\in\mathcal{N}(0, 1)$. We use the mean squared error~(MSE) to optimize the baseline model as shown in Eq:~\ref{eq:1}. To study the consistency of the results, 5-fold cross-validation is used on the IEMOCAP database while the transfer data sets are identical over the iterations and are used for model testing only. The models were trained over 50 epochs with a batch size of 64, and we used the Adam optimizer with the learning rate set to 1e-3. Additional methods used for reference are listed in the following section. 

\paragraph*{Reference methods:}
\label{sec:reference_methods}
In addition to comparing the performance of the DAE models with the proposed metric loss to the unsupervised DAE, we employ the following as reference methods to gauge the relative performance of the models developed in this work:
\vspace{-0.3cm}
\paragraph*{1.~}Since the DAE-unsupervised model is employed as a baseline to evaluate the effectiveness of the metric-loss, we compared its classification performance with similar methods from literature as presented in Table~\ref{tab:baseline}. 
\vspace{-0.3cm}
\paragraph*{2.}We implemented the support vector classifier, SVC, trained separately on all data sets towards the downstream task of classifying the input eGeMAPS features into target classes. The results are shown in Table~\ref{tab:classification}. We use the supervised SVC to study the upper-bound of the performance limits of the SER task. However, the work in this paper is focused on enabling reliable SER for languages with {\it few or no} labelled data, and supervised learning is therefore inapplicable within this scope.
\vspace{-0.3cm} 
\paragraph*{3.}We used the SUPERB: Speech processing Universal PERformance Benchmark for emotion classification on the transfer datasets~\cite{yang2021superb}. The method employs the HuBERT~(Hidden-Unit BERT), specifically hubert-large-ll60k model, as the base model to first extract speech representations~\cite{9585401}. The speech representations are then given as input to a linear downstream task to classify the speech signal into emotions. The base model is trained on English data sets and the linear downstream model for emotion classification is trained on the IEMOCAP dataset. We employ the model for classification only as the model is specifically trained for that. The results are shown in Table~\ref{tab:classification}. Relative to the complexity of the models developed in this paper~($<4\times10^2$ parameters), note that the considered model is highly complex~($>3\times10^8$ parameters). 
 \vspace{-0.3cm}
\paragraph*{4.}Lastly, besides comparing the correlation between the embedding and the labels for the proposed models and the unsupervised model, we employ supervision to train models with the proposed metric losses. In other words, we use the labels from the transfer data sets in the metric loss during training. We do this to obtain insights on the performance limits of the proposed method and the results are presented in Table~\ref{tab:correlation} as metric-act (supervised) and metric-val (supervised).   
\vspace{-0.2cm}
\paragraph*{Experiments for evaluation:}To investigate the efficacy of the proposed methods, in the following parts we investigate the quality of the latent embedding by using them as the input features for classification and correlation downstream tasks. Towards that, we evaluate the models in terms of \begin{enumerate*} \item the correlation coefficient, and \item the classification accuracy. \end{enumerate*}
\vspace{-0.3cm}
\paragraph*{1.~Correlation analysis}: We study the correlation between the latent dimensions and dimensional variables~(activation, valence). With the main focus to study the effectiveness of the proposed continuous metric learning method using the dimensional variables, we evaluate how large a proportion of the labels can be explained by the latent dimensions. Therefore, as described in Sec.~\ref{sec:OLS}, we compute the correlation coefficients between $\mathbf{z}$ and $l$, via the adjusted $R^2$. The mean and standard deviation over 5-folds for $R^2$ for the models are shown in Table~\ref{tab:correlation}; Note that the $R^2$, where the p-value of the F-statistic $>0.05$, are indicated by an asterisk. Models with larger $R^2$ are more effective in preserving the distance between samples in the embedding space with respect to the dimensional variables.
\vspace{-0.3cm}
\paragraph*{2.~Classification of emotion classes}:  We inspect the performance of the methods by classifying the speech samples into emotional categories using the support vector classifier~(SVC) with a linear kernel. For evaluation, we use balanced accuracy for the 4-class~(N-S-H-A) and 3-class~(N-S-A) scenarios to account for imbalanced classes. Furthermore, supervised training on eGeMAPS using SVC and the SUPERB model are employed as references, as described above. The results are presented in Table~\ref{tab:classification}.

\section{Results and discussion}
\begin{table}[!htb]

\resizebox{\columnwidth}{!}{%

\begin{tabular}{ |c||c|c|c|  }
 \hline
 Method & Features+Dataset & classes & Accuracy\\
 \hline
 GAN~\cite{latif2019unsupervised}   & eGeMAPS~\cite{eyben2015geneva}+EMO-DB    &2&   66\%~(UAR)\\
FLUDA~\cite{ahn2021cross}&   IS10~\cite{schuller2010interspeech}+IEMOCAP(+)  & 4   &50\%~(UA)\\
VAE+LSTM~\cite{latif2018variational}& LogMel+IEMOCAP & 4 & 56.08\%~(UA)\\
AE+LSTM~\cite{latif2018variational}& LogMel+IEMOCAP & 4 & 55.42\%~(UA)\\
Stacked-AE+BLSTM-RNN~\cite{ghosh2016representation}& COVAREP+IEMOCAP~\cite{degottex2014covarep} & 4 & 50.26\%~(UA)\\
DAE+Linear-SVM~\bf(baseline)& eGeMAPS+IEMOCAP & 4 & 52.09\%~(UA)\\
 \hline
\end{tabular}
}
 \caption{Performance in terms of unweighted accuracy~(UA) and unweighted average recall~(UAR) of the reference methods~(from cited papers) and the unsupervised baseline model developed in this paper.}
\label{tab:baseline}
 \end{table}
\begin{table*}[!htb]
    \resizebox{.99\textwidth}{!}{
    \begin{centering}
\begin{tabular}{lcccccccccc}
\toprule
\multirow{2}{*}{Method (DAE)}&\multicolumn{2}{c}{IEMOCAP}&\multicolumn{2}{c}{EMO-DB}&\multicolumn{2}{c}{CAFE}&\multicolumn{2}{c}{URDU}&\multicolumn{2}{c}{AESD}\\
 & $R^2$-Act & $R^2$-Val & $R^2$-Act& $R^2$-Val & $R^2$-Act & $R^2$-Val & $R^2$-Act & $R^2$-Val & $R^2$-Act & $R^2$-Val\\
\midrule
Metric-act (supervised) & NA & NA  & $0.38 \pm 0.05 $  & $0.16 \pm 0.04 $  & $0.62 \pm 0.01 $  & $0.16 \pm 0.01 $  & $0.34 \pm 0.05 $  & $0.15 \pm 0.04 $  & $0.44 \pm 0.03 $  & $0.18 \pm 0.01 $ \\
Metric-val (supervised) & NA & NA  & $0.45 \pm 0.03 $  & $0.21 \pm 0.03 $  & $0.44 \pm 0.05 $  & $0.29 \pm 0.06 $  & $0.32 \pm 0.06 $  & $0.16 \pm 0.04 $  & $0.4 \pm 0.06 $  & $0.17 \pm 0.03 $ \\
\hline
DAE-Unsupervised & $0.41 \pm 0.04 $  & $0.06 \pm 0.02 $  & $\mathbf{0.63 \pm 0.04 }$  & $0.05 \pm 0.04 $  & $0.41 \pm 0.03 $  & $0.14 \pm 0.02 $  & $0.28 \pm 0.05 $  & $0.14 \pm 0.03 $  & $0.3 \pm 0.01 $  & $-0.0 \pm 0.0 ^* $ \\
DAE-Metric-act & $\mathbf{0.49 \pm 0.02} $  & $0.05 \pm 0.01 $  & $\mathbf{0.63 \pm 0.04 }$  & $0.04 \pm 0.02 $  & $\mathbf{0.46 \pm 0.02} $  & $0.13 \pm 0.03 $  & $0.32 \pm 0.06 $  & $0.13 \pm 0.02 $  & $\mathbf{0.31 \pm 0.05 }$  & $-0.0 \pm 0.0 ^* $ \\
DAE-Metric-val & $0.39 \pm 0.03 $  & $\mathbf{0.11 \pm 0.01 }$  & $0.61 \pm 0.03 $  & $\mathbf{0.1 \pm 0.04 }$  & $0.43 \pm 0.02 $  & $\mathbf{0.15 \pm 0.01}$  & $\mathbf{0.38 \pm 0.01 }$  & $\mathbf{0.17 \pm 0.03 }$  & $0.27 \pm 0.03 $  & $0.01 \pm 0.01 ^* $ \\
\end{tabular}

\end{centering}
    }

    \caption{Adjusted squared correlation coefficient presenting the linear dependence of $l$ on $z$, the activation and valence labels for the three models. Mean and standard deviation over five folds are presented.}
        \label{tab:correlation}
\end{table*}

\begin{table*}[!htb]
    \resizebox{1.05\textwidth}{!}{

    \begin{centering}
\begin{tabular}{lcccccccccccc}
\toprule
\multirow{2}{*}{Method}&\multicolumn{2}{c}{IEMOCAP ($\mu\pm\sigma$)}&\multicolumn{2}{c}{EMO-DB ($\mu\pm\sigma$)}&\multicolumn{2}{c}{SAVEE ($\mu\pm\sigma$)}&\multicolumn{2}{c}{CAFE ($\mu\pm\sigma$)}&\multicolumn{2}{c}{URDU ($\mu\pm\sigma$)}&\multicolumn{2}{c}{AESD ($\mu\pm\sigma$)}\\
 & N-S-A & N-S-H-A & N-S-A & N-S-H-A & N-S-A & N-S-H-A & N-S-A & N-S-H-A & N-S-A & N-S-H-A & S-A & S-H-A\\
\midrule
SVC (supervised) & $0.65 \pm 0.02 $  & $0.65 \pm 0.02 $  & $0.89 \pm 0.03 $  & $0.68 \pm 0.03 $  & $0.74 \pm 0.03 $  & $0.68 \pm 0.05 $  & $0.66 \pm 0.03 $  & $0.51 \pm 0.03 $  & $0.89 \pm 0.03 $  & $0.82 \pm 0.02 $  & $0.94 \pm 0.03 $  & $0.7 \pm 0.06 $ \\
\hline
SUPERB~($>3\times10^8$) & $\mathbf{0.79}$  & $\mathbf{0.79 }$  & $0.57 $  & $\mathbf{0.66} $  & $\mathbf{0.7 }$  & $\mathbf{0.68 }$  & $0.39 $  & $\mathbf{0.51 }$  & $0.26 $  & $0.39 $  & $0.34 $  & $\mathbf{0.53 }$ \\
DAE-Unsupervised$^\dag$ & $0.51 \pm 0.02 $  & $0.51 \pm 0.02 $  & $0.72 \pm 0.06 $  & $0.56 \pm 0.05 $  & $0.59 \pm 0.02 $  & $0.49 \pm 0.02 $  & $0.43 \pm 0.0 $  & $0.32 \pm 0.01 $  & $0.51 \pm 0.05 $  & $0.38 \pm 0.03 $  & $0.4 \pm 0.05 $  & $0.22 \pm 0.03 $ \\
DAE-Metric-act$^\ddag$ & $0.54 \pm 0.02 $  & $0.54 \pm 0.01 $  & $0.74 \pm 0.04 $  & $0.57 \pm 0.04 $  & $0.58 \pm 0.02 $  & $0.46 \pm 0.03 $  & $\mathbf{0.46 \pm 0.04 }$  & $0.33 \pm 0.02 $  & $0.55 \pm 0.01 $  & $0.41 \pm 0.03 $  & $\mathbf{0.44 \pm 0.02 }$  & $0.27 \pm 0.02 $ \\
DAE-Metric-val$^\ddag$ & $0.54 \pm 0.01 $  & $0.54 \pm 0.02 $  & $\mathbf{0.78 \pm 0.03} $  & $0.61 \pm 0.03 $  & $0.61 \pm 0.05 $  & $0.49 \pm 0.02 $  & ${0.45 \pm 0.01 }$  & $0.34 \pm 0.02 $  & $\mathbf{0.6 \pm 0.02 }$  & $\mathbf{0.43 \pm 0.02 }$  & $0.42 \pm 0.02 $  & $0.25 \pm 0.02 $ \\
($<4\times10^2$ parameters)& & & & & & & & & & & &\\
\bottomrule
\end{tabular}
\end{centering}
}

\caption{Balanced classification accuracy for (a)~three emotion classes~(neutral, sad, anger) and (b)~four emotion classes~(neutral, sad, happy, anger) presented using mean and standard deviation~($\mu\pm\sigma$) computed over 5-fold cross validation. $\dag$ and $\ddag$ represents the baseline and proposed methods, respectively. Complexity of SUPERB and proposed models are presented in parentheses.}
\label{tab:classification}
\end{table*}

In Table.~\ref{tab:baseline}, we list current models similar to the DAE-unsupervised baseline in terms of the architecture, network size and input-output format. We observe that the  performance of the developed unsupervised model is comparable to state of the art. Therefore, we consider the DAE+Linear-SVM model a reasonable baseline to gauge the performance of the proposed metric-loss in this paper. 
\vspace{-0.3cm}
\paragraph*{Correlation analysis:}
From Table~\ref{tab:correlation}, we observe that the mean and standard deviation of $R^2$ is consistently higher for the proposed methods relative to the unsupervised baseline, over the transfer data sets. However, DAE-unsupervised is as good as DAE-metric-act in terms of the correlation between $\mathbf{z}$ and the activation for EMO-DB. With supervision, as anticipated metric-act and metric-val models seem to have a higher $R^2$ relative to the unsupervised and semi-supervised models, specifically for the activation variable. \\
To summarize the observations, the proposed methods show higher adjusted $R^2$ between the modeled $\mathbf{z}$ and the label considered. However, we observe that $R^2$ corresponding to the valence variable is lower than the activation variable. An unaccounted non-linear relation between valence variable and the latent space could be a reason for the observation. Nevertheless, both metric-act and metric-val seem to effectively order and arrange data samples in the latent space, relative to the activation label. This is also evident from Fig.~\ref{fig:latent_emb}, wherein we can observe that the samples in the latent space are better distributed for metric-act and metric-val then for the unsupervised model.  
\vspace{-0.3cm}
\paragraph*{Classification of emotion classes:}
 We observe that for both 3- and 4-class scenarios, the proposed methods have higher accuracy than the baseline unsupervised method. Furthermore, while SUPERB outperforms all the methods for IEMOCAP and SAVEE, it seems to have lower accuracy than the baseline and the proposed models for the remaining data sets, specifically for the 3-class classification task. Lastly, although classification accuracy of the supervised SVC is superior and implies that there is class-discriminating information in the data sets, how much of that information corresponds to the paralinguistic aspects of speech and emotion is worth investigating.\\
The balanced accuracy scores~(Table~\ref{tab:classification}) for metric-val and metric-act is consistently better than the unsupervised baseline over the transfer datasets, indicating that incorporating a distance preservation metric in the loss function aids in shaping the distribution of features in the latent space that is more consistent over languages. It is also interesting to note that while metric-val did not show a large difference in $R^2$ with respect to the baseline~(Table~\ref{table:1}), in terms of classification accuracy it is often better than metric-act. Furthermore, the proposed models are unable to effectively differentiate between anger and happy, whereby its performance drops in 4-class classification. A similar trend was observed for the reference SUPERB model~(lower 3-class classification accuracy), despite its much higher complexity. This indicates that more complex models do not necessary lead to learning meaningful representations of the more subtle aspects of speech. 
\vspace{-0.3cm}
\paragraph*{Effectiveness of metric loss function:} From Table~\ref{table:1} we observe that metric-act, i.e., the DAE trained with semi-supervision from the activation labels of the IEMOCAP dataset, shows the highest adjusted $R^2$ value between $\mathbf{l_d}$ and $\mathbf{z_d}$, $\mathbf{l_d}$ corresponding to the activation labels. This suggests that the proposed formulation of the continuous metric loss~(Eq.~\ref{eq:met}) is effective when activation labels are employed for semi-supervision. In contrast, for metric-val wherein $\mathbf{l_d}$ corresponds to valence, the $R^2$ value is similar to the baseline. A potential reason for this could be that the relation between valence and the embedding is inherently non-linear. This implies that different approaches are necessary to model valence and activation variables. Investigating and including aspects of the true relation between valence, activation and the embedding will be addressed in future work. In conclusion, we can state that using the dimensional variables, activation and valence, to learn emotion representations yield embeddings that are more transferable to unseen datasets and new languages. Furthermore, the proposed continuous metric loss with semi-supervision enables us to incorporate information on the dimensional variables within the model, hence aiding transferability. 
\vspace{-0.3cm}
\section{Conclusion}
\vspace{-0.3cm}
In this work, we proposed a method for continuous metric learning, such that the difference between two feature points is continuous. We apply the method for speech emotion recognition, wherein we model a DAE that is semi-supervised using the activation and valence labels and the continuous metric loss. Our results show that the embedding from the proposed method is generally more consistent and thereby more transferable to different languages. The proposed methods are evaluated in terms of classification performance, and the proposed models outperform the baseline method on all the transfer datasets. Furthermore, to investigate the correspondence of the latent space to activation and valence variables, we compute the adjusted $R^2$ that indicates how much variation in the labels can be explained by a linear combination of the latent space. While the $R^2$ with respect to valence is generally lower than that for activation, between the methods, the proposed models outperform the baseline method for the datasets. Addressing the lack of labels corresponding to activation and valence variables in the transfer dataset, the annotated transfer dataset is the second contribution of this work.  

\bibliographystyle{abbrv}
\bibliography{references}
\balance
\end{document}